\documentclass[12pt,twoside,english]{article}
\usepackage[T1]{fontenc}
\usepackage[latin1]{inputenc}
\usepackage{geometry}
\usepackage{babel}
\usepackage{graphics}

\begin{document}
\begin{titlepage}
\setcounter{page}{1}
\rightline{}
\vspace{4.0cm}
\begin{center}
{\Large \bf Leading Twist Amplitudes for Exclusive Neutrino Interactions}
\vspace{.17in}
{\Large \bf in the Deeply Virtual Limit} 
\vspace{4cm}

{\large Claudio Corian\`{o} and  Marco Guzzi}\\ 
\vspace{.12in}
\vspace{.12in}
{\it  Dipartimento di Fisica, Universit\`{a} di Lecce \\
and INFN Sezione di Lecce \\ Via Arnesano 73100 Lecce, Italy}\\

\end{center}
\vspace{3cm}

\begin{abstract}
Neutrino scattering on nucleons in the regime of deeply 
virtual kinematics is studied both in the charged and the neutral 
electroweak sectors 
using a formalism developed by Bl\"umlein, Robaschik, Geyer and 
Collaborators for the analysis
of the Virtual Compton amplitude in the generalized Bjorken region. 
We discuss the structure of the leading twist amplitudes of the process.

\end{abstract}
\smallskip
\end{titlepage}
\setcounter{footnote}{0}


\def\beq{\begin{equation}}
\def\eeq{\end{equation}}
\def\beqn{\begin{eqnarray}}
\def\eeqn{\end{eqnarray}}
\def\ba{\begin{eqnarray}}
\def\ea{\end{eqnarray}}
\def\ie{{\it i.e.}}
\def\eg{{\it e.g.}}
\def\half{{\textstyle{1\over 2}}}
\def\nicefrac#1#2{\hbox{${#1\over #2}$}}
\def\third{{\textstyle {1\over3}}}
\def\quarter{{\textstyle {1\over4}}}
\def\m{{\tt -}}
\def\p{{\tt +}}
\def\slash#1{#1\hskip-6pt/\hskip6pt}
\def\slk{\slash{k}}
\def\GeV{\,{\rm GeV}}
\def\TeV{\,{\rm TeV}}
\def\y{\,{\rm y}}
\def\ds{\slash}
\def\l{\langle}
\def\r{\rangle}
\def\xprime{x^{\prime}}
\def\xprimetwo{x^{\prime\prime}}
\def\zprime{z^{\prime}}
\def\xprimbar{\overline{x}^\prime}
\def\xprim2bar{\overline{x}^{\prime\prime}}
\def\ptbold{\mbox{\boldmath$p$}_T}
\def\ktbold{\mbox{\boldmath$k$}_T}
\def\ktboldbar{\mbox{\boldmath$\overline{k}$}_T}
\def\beq{\begin{equation}}
\def\eeq{\end{equation}}
\def\tr{{\bf tr}}
\def\P{P^\mu}
\def\Pb{\overline{P}^\mu}
\def\BOX#1#2#3#4#5{\hskip#1mm\raisebox{#2mm}[#3mm][#4mm]{$#5$}}
\def\VBOX#1#2{\vbox{\hbox{#1}\hbox{#2}}}

\setcounter{footnote}{0}
\newcommand{\beqa}{\begin{eqnarray}}
\newcommand{\eeqa}{\end{eqnarray}}
\newcommand{\eps}{\epsilon}
\pagestyle{plain}
\setcounter{page}{1}

\section{Introduction and Motivations} 
Exclusive processes mediated by the weak force are
an area of investigation which may gather a wide interest 
in the forthcoming years due to the various experimental proposals to detect 
neutrino oscillations at intermediate energy using neutrino factories and superbeams 
\cite{Marciano1}. These proposals require a study of the neutrino-nucleon interaction 
over a wide range of energy starting from the elastic/quasi-elastic domain 
up to the deep inelastic scattering (DIS) region (see \cite{Winter},\cite{Mangano},\cite{Morfin},\cite{nuint},\cite{Quigg} for an overall overview).     
However, the discussion of the neutrino nucleon interaction has, so far, 
been confined either to the DIS region or to the form factor/nucleon 
resonance region, while the intermediate energy region, at this time, 
remains unexplored also theoretically. Clearly, to achieve a ``continuos'' description of the underlying strong interaction 
dynamics, from the resonant to the perturbative regime, will require considerable effort, since it is experimentally and 
theoretically difficult to disentangle a perturbative from a non-perturbative dynamics at intermediate energy, which 
appear to be superimposed. This is best exemplified - at least in the case of electromagnetic processes, such as Compton scattering - in the dependence of the intermediate energy description on the momentum transfer \cite{CL}.  
In this respect, the interaction of neutrinos with the constituents of the nucleon is no different, once the partonic structure of the target is resolved. From our viewpoint, 
the presence of such a gap in our knowledge well justifies any 
attempt to improve the current situation. 

Together with Amore, 
we have pointed out \cite{ACG} 
that exclusive processes of DVCS-type (Deeply Virtual Compton Scattering) 
could be relevant also in the theoretical study of the exclusive neutrino/nucleon interaction.  
Thanks to the presence 
of an on-shell photon emitted in the final state, this particle 
could be tagged together with the recoiling nucleon 
in a large underground detector in order to trigger on the process and 
exclude contamination from other backgrounds.   
With these motivations, 
a study of the $\nu N\to \nu N \gamma$ process has been performed 
in \cite{ACG}. The process is mediated by a neutral current and 
is particularly clean since there is no 
Bethe-Heitler contribution. It has been termed 
{\em Deeply Virtual Neutrino Scattering} or DVNS and requires in its partonic 
description the electroweak 
analogue of the ``non-forward parton distributions'', previously 
introduced in the study of DVCS.

 In this work we extend that analysis and provide, in part, 
a generalization of those results to the charged current case. 
Our treatment, here, is purposely short. 
The method that we use for the study of the charged processes is based on the 
formalism of the non-local operator product expansion and the technique of the 
harmonic polynomials, which allows to classify the various contributions 
to the interaction in terms of operators of a definite geometrical twist 
\cite{Lazar}. 
We present here a classification of the leading twist amplitudes 
of the charged process while a detailed phenomenological analysis useful 
for future experimental searches will be given elsewhere.

\section{The Generalized Bjorken Region and DVCS}
\begin{figure}[t]
{\par\centering \resizebox*{12cm}{!}{\includegraphics{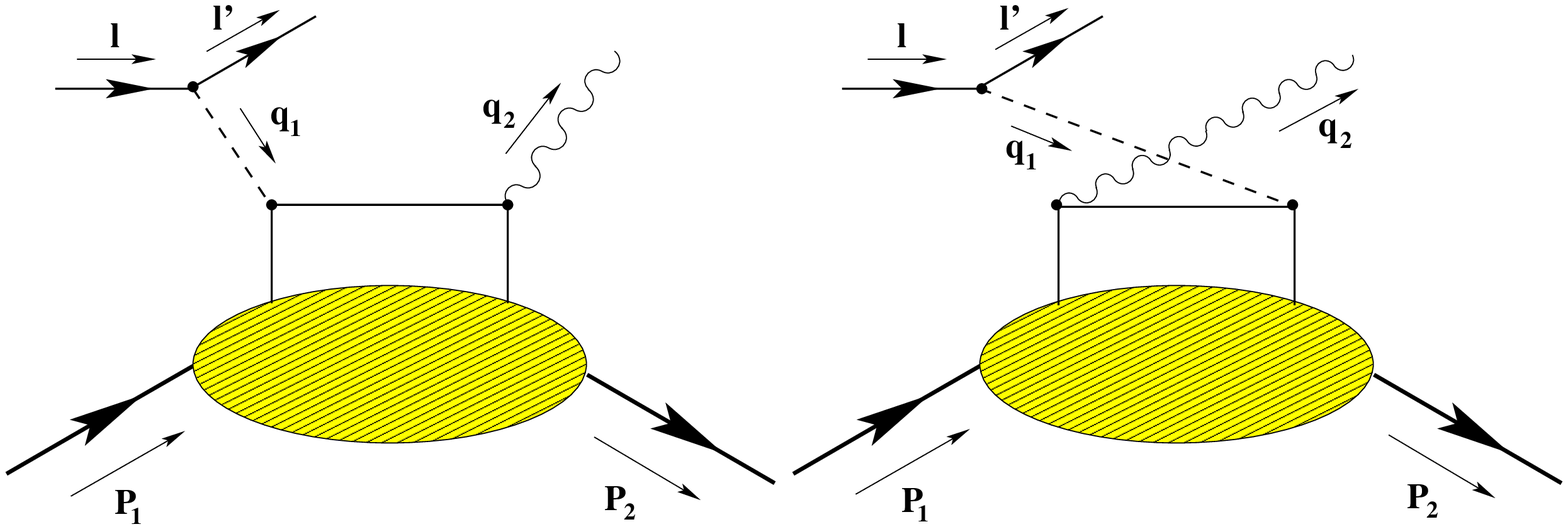}} \par}
\caption{Leading hand-bag diagrams for the process}
\label{DVCS_1.eps}
\end{figure}

Fig.~\ref{DVCS_1.eps} illustrates the process that we are going to study, 
where a neutrino of momentum $l$ scatters off a nucleon of momentum $P_1$ 
via a neutral or a charged current interaction; 
from the final state a photon and a nucleon emerge, of momenta $q_2$ and $P_2$ respectively, while the momenta of the final lepton is $l'$. 
We recall that Compton scattering has been investigated in the near past by several groups, since the original works \cite {Ji1, Rad3, Rad1}.
A previous study of the Virtual Compton process in the generalized Bjorken region, of which DVCS 
is just a particular case, can be found in \cite{Geyer}.
From the hadronic side, the description of the interaction proceeds 
via new constructs of the parton model termed 
{ \em generalized parton distributions} (GPD) or also 
{\em non-forward parton distributions}. The kinematics for the study of GPD's is characterized by a deep virtuality of the exchanged photon in the initial interaction 
($\nu +p\to \nu +p +\gamma$) ( $ Q^2 \approx $  2 GeV$^2$), with the final state photon kept on-shell; large energy of the hadronic system ($W^2 > 6$ GeV$^2$) above the resonance domain and small momentum transfers $|t| < 1$ GeV$^2$. 
In the electroweak case, photon emission can occur from the final state 
electron (in the case of charged current interactions) and provides an additional contribution to the virtual Compton 
amplitude. We choose symmetric defining momenta and use as independent variables the average of the hadron and gauge bosons momenta 

\beq
P_{1,2}= \bar{P} \mp \frac{\Delta}{2}\,\,\,\,\,\,\,\, q_{1,2}= q \pm \frac{\Delta}{2},
\eeq

with $\Delta=P_2-P_1$ being the momentum transfer. Clearly 

\beq
\bar{P}\cdot \Delta=0,\,\,\,\,\, \bar{P}^2=M^2 - \frac{\Delta^2}{4}
\eeq

and $M$ is the nucleon mass. There are two scaling variables which are identified in the process, since 3 scalar products  can grow large in the generalized Bjorken limit: $q^2$, $\Delta\cdot q$, $\bar{P}\cdot q$. 

The momentum transfer $\Delta^2$ is a small parameter in the process. Momentum asymmetries between the initial and the final state nucleon are measured by two scaling parameters, $\xi$ and $\eta$, related to ratios of the former invariants 

\ba
\xi=-\frac{q^2}{2 \bar{P}\cdot q}&& \eta=\frac{\Delta\cdot q}{2 \bar{P}\cdot q}
\ea

where $\xi$ is a variable of Bjorken type, expressed in terms of average momenta rather than nucleon and  gauge bosons momenta.  
The standard Bjorken variable $x= - q_1^2/( 2 P_1\cdot q_1)$ is trivially related to $\xi$ in the $t=0$ limit and in the DVCS case $\eta=-\xi$. 

Notice also that the parameter $\xi$ measures the ratio between the plus component of the momentum transfer and the average momentum. 

$\xi$, therefore, parametrizes the large component of the momentum transfer $\Delta$, which can be generically described as 

\beq
\label{delta}
\Delta= -2 \xi \bar{P} -{\Delta}_{\perp}
\eeq

where all the components of ${\Delta}_{\perp}$ are $O\left(\sqrt{|\Delta^2|}\right)$.

\section{Bethe-Heitler Contributions}
Prior to embark on the discussion of the virtual Compton contribution, 
we quote the result for the Bethe-Heitler (BH) subprocess, which makes its first 
appearance in the charged 
current case, since a real photon can be radiated off the 
leg of the final state lepton.
The amplitude of the BH contribution for a $W^+$ exchange is as follows
\ba   
&&T_{BH}^{W^+}=-|e| \frac{g}{2\sqrt{2}}\frac{g}{\sqrt{2}}\overline{u}(l')\left[\gamma^{\mu}\frac{(\ds{l}-\ds{\Delta})}{(l-\Delta)^2 +i\eps}\gamma^{\nu}(1-\gamma^5)\right]u(l)\frac{D^{\nu\delta}(q_1)}{\Delta^2 -M_{W}^{2}+i\eps}\eps^{*}_{\mu}(q_2)\times\nonumber\\
&&\hspace{2 cm}\overline{U}(P_2)\left[\left(F_{1}^{u}(\Delta^2)-F_{1}^{d}(\Delta^2)\right)\gamma^{\delta}+\left(F_{2}^{u}(\Delta^2)-F_{2}^{d}(\Delta^2)\right)i\frac{\sigma^{\delta\alpha}\Delta_{\alpha}}{2 M}\right]U(P_1),\nonumber\\ 
\ea
where $\eps$ is the polarization vector of the photon
and
\ba
&&T_{BH}^{W^-}=|e| \frac{g}{2\sqrt{2}}\frac{g}{\sqrt{2}}\overline{v}(l)\left[\gamma^{\mu}\frac{(\ds{l}-\ds{\Delta})}{(l-\Delta)^2 +i\eps}\gamma^{\nu}(1-\gamma^5)\right]v(l')
\frac{D^{\nu\delta}(q_1)}{\Delta^2 -M_{W}^{2}+i\eps}\eps^{*}_{\mu}(q_2)\times\nonumber\\
&&\hspace{2 cm}\overline{U}(P_2)\left[\left(F_{1}^{u}(\Delta^2)-F_{1}^{d}(\Delta^2)\right)\gamma^{\delta}+\left(F_{2}^{u}(\Delta^2)-F_{2}^{d}(\Delta^2)\right)i\frac{\sigma^{\delta\alpha}\Delta_{\alpha}}{2 M}\right]U(P_1)\,\nonumber\\
\ea
for the $W^-$ case, with 
$D^{\nu\delta}(q_1)/{(\Delta^2 -M_{W}^{2}+i\eps)}$ being the propagator of the W's and 
$F_{1,2}$ the usual nucleon form factors (see also 
\cite{ACG}).
\section{Structure of the Compton amplitude for charged and neutral currents}  
Moving to the Compton amplitude for charged and neutral currents, 
this can be expressed in terms of the correlator of currents
\ba
T_{\mu \nu}=i\int d^{4}x e^{i qx} \langle P_2|T\left(J_{\nu}^{\gamma}(x/2) J_{\mu}^{W^{\pm},Z_{0}}(-x/2)\right)|P_1\rangle\,, 
\ea
where for the charged and neutral currents we have the following expressions
\ba
&&J^{\mu Z_{0}}(-x/2)=\frac{g}{2 \cos{\theta_W}}\overline{\psi}_{u}(-x/2)\gamma^{\mu}(g^{Z}_{u V}+g^{Z}_{u A}\gamma^5)\psi_{u}(-x/2)+\overline{\psi}_{d}(-x/2)\gamma^{\mu}(g^{Z}_{d V}+g^{Z}_{d A}\gamma^5)\psi_{d}(-x/2),\,\nonumber\\
&&J^{\mu W^{+}}(-x/2)=\frac{g}{2 \sqrt{2}}\overline{\psi}_{u}(-x/2)\gamma^{\mu}(1-\gamma^5)U^{*}_{u d}\psi_{d}(-x/2),\,\nonumber\\
&&J^{\mu W^{-}}(-x/2)=\frac{g}{2 \sqrt{2}}\overline{\psi}_{d}(-x/2)\gamma^{\mu}(1-\gamma^5)U_{d u}\psi_{u}(-x/2),\,\nonumber\\
&&J^{\nu, \gamma}(x/2)=\overline{\psi}_{d}(x/2)\gamma^{\nu}\left(-\frac{1}{3}e\right)\psi_{d}(x/2) +\overline{\psi}_{u}(x/2)\gamma^{\nu}\left(\frac{2}{3}e\right)\psi_{u}(x/2)\,.
\ea
Here we have chosen a simple representation of 
the flavour mixing matrix $U^{*}_{u d}=U_{u d}=U_{d u}=\cos{\theta_{C}}$, where $\theta_{C}$ is the Cabibbo angle. 

The coefficients $g_V^{Z}$ and $g_A^{Z}$ are 
\ba
&&g^{Z}_{u V}=\frac{1}{2} + \frac{4}{3} \sin^{2}\theta_{W}\hspace{1.5 cm}g^{Z}_{u A}=-\frac{1}{2}\nonumber\\
&&g^{Z}_{d V}=-\frac{1}{2} + \frac{2}{3} \sin^{2}\theta_{W}\hspace{1.5 cm}g^{Z}_{d A}=\frac{1}{2}\,,
\ea
and 
\ba
g_{u}=\frac{2}{3},&&g_{d}=\frac{1}{3}
\ea
are the absolute 
values of the charges of the up and down quarks in units 
of the electron charge. A short computation gives
\ba
&&\langle P_2|T\left(J_{\nu}^{\gamma}(x/2) J_{\mu}^{Z_{0}}(-x/2)\right)|P_1\rangle=\nonumber\\
&&\langle P_2|\overline{\psi}_{u}(x/2)g_u \gamma_{\nu}S(x)\gamma_{\mu}(g^{Z}_{u V}+g^{Z}_{u A}\gamma^5)\psi_{u}(-x/2)-\nonumber\\
&&\hspace{0.8 cm}\overline{\psi}_{d}(x/2)g_d \gamma_{\nu}S(x)\gamma_{\mu}(g^{Z}_{d V}+g^{Z}_{d A}\gamma^5)\psi_{d}(-x/2)+\nonumber\\
&&\hspace{0.8 cm}\overline{\psi}_{u}(x/2)\gamma_{\mu}(g^{Z}_{u V}+g^{Z}_{u A}\gamma^5)S(-x) g_u \gamma_{\nu}\psi_{u}(x/2)-\nonumber\\
&&\hspace{0.8 cm}\overline{\psi}_{d}(x/2)\gamma_{\mu}(g^{Z}_{d V}+g^{Z}_{d A}\gamma^5)S(-x) g_d \gamma_{\nu}\psi_{d}(x/2)|P_1 \rangle,\, \nonumber\\
\\
&&\langle P_2|T\left(J_{\nu}^{\gamma}(x/2) J_{\mu}^{W^{+}}(-x/2)\right)|P_1\rangle=\nonumber\\
&&\langle P_2|\overline{\psi}_{u}(-x/2)\gamma_{\mu}(1-\gamma^5)U_{u d}S(-x)\gamma_{\nu}\left(-g_{d}\right)\psi_{d}(x/2)+\nonumber\\
&&\hspace{0.8 cm}\overline{\psi}_{u}(x/2)\gamma_{\nu}\left(g_{u}\right)S(x)\gamma_{\mu}(1-\gamma^5)U_{u d}\psi_{d}(-x/2) |P_1\rangle,\,\nonumber\\
\\
&&\langle P_2|T\left(J_{\nu}^{\gamma}(x/2) J_{\mu}^{W^{-}}(-x/2)\right)|P_1\rangle=\nonumber\\
&&\langle P_2|-\overline{\psi}_{d}(x/2)g_d\gamma_{\nu}S(x)\gamma_{\mu}(1-\gamma^5)U_{d u}\psi_{u}(-x/2)+\nonumber\\
&&\hspace{0.8 cm}\overline{\psi}_{d}(-x/2)\gamma_{\mu}(1-\gamma^5)S(-x)U_{d u}\psi_{u}(x/2) |P_1\rangle\,,
\ea
where all the factors $g/{2\sqrt{2}}$ and $g/2\cos{\theta_{W}}$, 
for semplicity, have been suppressed  and we have defined 

\ba
S^{u}(x)=S^{d}(x)\approx \frac{i \rlap/{x}}{2\pi^2(x^2-i\eps)^2}.
\ea

Using the following identities 
\ba
&&\gamma_{\mu}\gamma_{\alpha}\gamma_{\nu}=S_{\mu\alpha\nu\beta}\gamma^{\beta}+i\eps_{\mu\alpha\nu\beta}\gamma^{5}\gamma^{\beta},\nonumber\\
&&\gamma_{\mu}\gamma_{\alpha}\gamma_{\nu}\gamma^{5}=S_{\mu\alpha\nu\beta}\gamma^{\beta}\gamma^{5}-i\eps_{\mu\alpha\nu\beta}\gamma^{\beta},\nonumber\\
&&S_{\mu\alpha\nu\beta}=\left(g_{\mu\alpha}g_{\nu\beta}+g_{\nu\alpha}g_{\mu\beta}-g_{\mu\nu}g_{\alpha\beta} \right),
\ea
we rewrite the correlators as 
\ba
&&T_{\mu\nu}^{Z_{0}}=i\int d^{4}x\frac{e^{i q x} x^{\alpha}}{2\pi^2(x^2 -i\eps)^2}
\langle P_2|\left[g_u g_{u V}\left(S_{\mu\alpha\nu\beta}O^{\beta}_{u} -i\eps_{\mu\alpha\nu\beta}O^{5 \beta}_{u}\right)-g_u g_{u A}\left(S_{\mu\alpha\nu\beta}\tilde{O}^{5 \beta}_{u}-i\eps_{\mu\alpha\nu\beta}\tilde{O}^{\beta}_{u}\right)\right.\nonumber\\
&&\hspace{5.5 cm}\left.-g_d g_{d V}\left(S_{\mu\alpha\nu\beta}O^{\beta}_{d}-i\eps_{\mu\alpha\nu\beta}O^{5 \beta}_{d}\right)+g_d g_{d A}\left(S_{\mu\alpha\nu\beta}\tilde{O}^{5 \beta}_{d} -i\eps_{\mu\alpha\nu\beta}\tilde{O}^{\beta}_{d}\right)\right]|P_1\rangle,\,\nonumber\\
\\
&&T^{W^{+}}_{\mu \nu}=i\int d^4 x \frac{e^{iqx}x^{\alpha}U_{u d}}{2\pi^2(x^2-i\eps)^2}\langle P_2|\left[i S_{\mu\alpha\nu\beta}\left(\tilde{O}_{u d}^{\beta}+O_{u d}^{5 \beta}\right)+\eps_{\mu\alpha\nu\beta}\left(O_{u d}^{\beta}+\tilde{O}_{u d}^{5 \beta}\right)\right] |P_1\rangle, \nonumber\\
\\
&&T^{W^{-}}_{\mu \nu}=i\int d^4 x \frac{e^{iqx}x^{\alpha}U_{d u}}{2\pi^2(x^2-i\eps)^2}\langle P_2|\left[-i S_{\mu\alpha\nu\beta}\left(\tilde{O}_{d u}^{\beta}+O_{d u}^{5 \beta}\right)-\eps_{\mu\alpha\nu\beta}\left(O_{d u}^{\beta}+\tilde{O}_{d u}^{5 \beta}\right)\right] |P_1\rangle\,. \nonumber\\
\ea
We have suppressed the $x$-dependence of the operators in the former equations. The relevant operators are denoted by 
\ba
&&\tilde{O}_{a}^{\beta}(x/2,-x/2)=\overline{\psi}_{a}(x/2)\gamma^{\beta}\psi_{a}(-x/2)+\overline{\psi}_{a}(-x/2)\gamma^{\beta}\psi_{a}(x/2),\nonumber\\
&&\tilde{O}_{a}^{5 \beta}(x/2,-x/2)=\overline{\psi}_{a}(x/2)\gamma^{5}\gamma^{\beta}\psi_{a}(-x/2)-\overline{\psi}_{a}(-x/2)\gamma^{5}\gamma^{\beta}\psi_{a}(x/2),\nonumber\\
&&O_{a}^{\beta}(x/2,-x/2)=\overline{\psi}_{a}(x/2)\gamma^{\beta}\psi_{a}(-x/2)-\overline{\psi}_{a}(-x/2)\gamma^{\beta}\psi_{a}(x/2),\nonumber\\
&&O_{a}^{5 \beta}(x/2,-x/2)=\overline{\psi}_{a}(x/2)\gamma^{5}\gamma^{\beta}\psi_{a}(-x/2)+\overline{\psi}_{a}(-x/2)\gamma^{5}\gamma^{\beta}\psi_{a}(x/2),\,\nonumber\\
\\
&&\tilde{O}_{u d}^{\beta}(x/2,-x/2)=g_u\overline{\psi}_{u}(x/2)\gamma^{\beta}\psi_{d}(-x/2)+g_d\overline{\psi}_{u}(-x/2)\gamma^{\beta}\psi_{d}(x/2),\nonumber\\
&&\tilde{O}_{u d}^{5 \beta}(x/2,-x/2)=g_u\overline{\psi}_{u}(x/2)\gamma^{5}\gamma^{\beta}\psi_{d}(-x/2)-g_d\overline{\psi}_{u}(-x/2)\gamma^{5}\gamma^{\beta}\psi_{d}(x/2),\nonumber\\
&&O_{u d}^{\beta}(x/2,-x/2)=g_u\overline{\psi}_{u}(x/2)\gamma^{\beta}\psi_{d}(-x/2)-g_d\overline{\psi}_{u}(-x/2)\gamma^{\beta}\psi_{d}(x/2),\nonumber\\
&&O_{u d}^{5 \beta}(x/2,-x/2)=g_u\overline{\psi}_{u}(x/2)\gamma^{5}\gamma^{\beta}\psi_{d}(-x/2)+g_d\overline{\psi}_{u}(-x/2)\gamma^{5}\gamma^{\beta}\psi_{d}(x/2),\,\nonumber\\
\ea
and similar ones with $u \leftrightarrow d$ interchanged.

We use isospin symmetry to relate flavour nondiagonal operators 
$(\hat{O}_{f f'})$ to flavour diagonal ones $(\hat{O}_{f f})$

\ba
\langle p|\hat{O}^{ud}(x)|n\rangle&=&\langle p|\hat{O}^{ud}(x)\tau^{-}|n\rangle=\langle p|\left[\hat{O}^{ud}(x),\tau^{-}\right]|n\rangle=
\langle p|\hat{O}^{uu}(x)|p\rangle-\langle p|\hat{O}^{dd}(x)|p\rangle\,,\nonumber\\
&&\langle p|\hat{O}^{ud}(x)|n\rangle=\langle n|\hat{O}^{dd}(x)|n\rangle-\langle n|\hat{O}^{uu}(x)|n\rangle\,,\nonumber\\
&&\langle n|\hat{O}^{du}(x)|p\rangle=\langle p|\hat{O}^{uu}(x)|p\rangle-\langle p|\hat{O}^{dd}(x)|p\rangle\,,\nonumber\\
&&\langle n|\hat{O}^{du}(x)|p\rangle=\langle n|\hat{O}^{dd}(x)|n\rangle-\langle n|\hat{O}^{uu}(x)|n\rangle,\,
\ea
where \ba
\tau^{\pm}=\tau^{x}\pm\tau^{y}
\ea
are isospin raising/lowering operators expressed in terms of Pauli matrices.  

\section{Parameterization of nonforward matrix elements}
The extraction of the leading twist contribution to the handbag diagram 
is performed using the geometrical twist expansion, as developed in 
\cite{blum&rob,lazar1,lazar2}, adapted to our case. We set 
the twist-2 expansions on the light cone (with $x^2=0$) 
and we choose the light-cone gauge to remove the gauge link

\ba
&&\langle P_2|\overline{\psi}_{a}(-k x)\gamma^{\mu}\psi_{a}(k x)|P_1\rangle^{tw.2}=\nonumber\\
&&\int Dz e^{-ik(x\cdot P_z)}F^{a (\nu)}(z_{1},z_{2},P_{i}\cdot P_{j}x^2,P_{i}\cdot P_{j})\overline{U}(P_2)\left[\gamma^{\mu}-i k {P_{z}}^{\mu}\ds{x}\right]U(P_1)+\nonumber\\
&&\int Dz e^{-ik(x\cdot P_z)}G^{a (\nu)}(z_{1},z_{2},P_{i}\cdot P_{j}x^2,P_{i}\cdot P_{j})\overline{U}(P_2)\left[\frac{\left(i\sigma^{\mu \alpha}\Delta_{\alpha}\right)}{M} -ik{P_{z}}^{\mu}\frac{\left(i\sigma^{\alpha \beta}x_{\alpha}\Delta_{\beta}\right)}{M}\right]U(P_1)\,,\nonumber\\
\ea 
with $0< k <1$ a scalar parameter, with
\ba
P_z=P_1 z_1 + P_2 z_2,\, 
\ea
and 

\ba
&&\langle P_2|\overline{\psi}_{a}(-k x)\gamma^{5}\gamma^{\mu}\psi_{a}(k x)|P_1\rangle^{tw.2}=\nonumber\\
&&\int Dz e^{-ik(x\cdot P_z)}F^{5 a (\nu)}(z_{1},z_{2},P_{i}\cdot P_{j}x^2,P_{i}\cdot P_{j})\overline{U}(P_2)\left[\gamma^{5}\gamma^{\mu}-i k {P_{z}}^{\mu}\gamma^{5}\ds{x}\right]U(P_1)+\nonumber\\
&&\int Dz e^{-ik(x\cdot P_z)}G^{5 a(\nu)}(z_{1},z_{2},P_{i}\cdot P_{j}x^2,P_{i}\cdot P_{j})\overline{U}(P_2)\gamma^{5}\left[\frac{\left(i\sigma^{\mu \alpha}\Delta_{\alpha}\right)}{M} -ik{P_{z}}^{\mu}\frac{\left(i\sigma^{\alpha \beta}x_{\alpha}\Delta_{\beta}\right)}{M}\right]U(P_1).\,\nonumber\\
\ea
The index $(\nu)$ in the expressions of the distribution functions $F, G$ has been introduced in order to distinguish 
them from the parameterization given in \cite{blum&rob,Geyer}, which are related to linear combinations of electromagnetic 
correlators. In the expressions above $a$ is a flavour index and we have introduced both a vector (Dirac) and a Pauli-type 
form factor contribution with nucleon wave functions (U(P)). 
The product $P_i\cdot P_j$ denotes all the possible products of the two momenta $P_1$ and $P_2$,
and the measure of integration is defined by \cite{blum&rob}
\ba
Dz=\frac{1}{2} d z_1 d z_2\,\theta(1-z_1)\,\theta(1+z_1)\,\theta(1-z_2)\,
\theta(1+z_2).\,
\ea

In our parameterization of the correlators we are omitting the so called ``trace-terms'' (see ref.~\cite{Geyer}), since these terms vanish on shell.
In order to arrive at a partonic interpretation  one introduces 
variables $z_{+}$ and $z_{-}$ conjugated to $2 \bar{P}$ and $\Delta$ and defined as
\ba
&&z_{+}=1/2(z_1 +z_2),\nonumber\\
&&z_{-}=1/2(z_2 -z_1),\nonumber\\
&&Dz=dz_{+} dz_{-}\theta(1+z_+ + z_-)\theta(1+z_+ - z_-)\theta(1-z_+ + z_-)\theta(1-z_+ - z_-).\,
\ea
In terms of these new variables 
$P_z=2\bar{P}z_{+}+\Delta z_{-}$, which will be used below.

At this stage, we can proceed to calculate the hadronic tensor 
by performing the $x$-space integrations. This will be illustrated in the 
case of the $W^+$ current, the others being similar. We define

\ba
&&\int dx^4 \frac{e^{iq x}x^{\alpha}}{2\pi^2\left(x^2-i\eps\right)^2}\langle P_2|S_{\mu\alpha\nu\beta}\tilde{O}^{a\beta}|P_1\rangle=\tilde{S}^{a}_{\mu\nu},\nonumber\\
&&\int dx^4 \frac{e^{iq x}x^{\alpha}}{2\pi^2\left(x^2-i\eps\right)^2}\langle P_2|S_{\mu\alpha\nu\beta} O^{5 a\beta}|P_1\rangle=S^{5 a}_{\mu\nu},\nonumber\\
&&\int dx^4 \frac{e^{iq x}x^{\alpha}}{2\pi^2\left(x^2-i\eps\right)^2}\langle P_2|\eps_{\mu\alpha\nu\beta} O^{a\beta}|P_1\rangle={\bf \varepsilon}^{a}_{\mu\nu},\nonumber\\
&&\int dx^4 \frac{e^{iq x}x^{\alpha}}{2\pi^2\left(x^2-i\eps\right)^2}\langle P_2|\eps_{\mu\alpha\nu\beta} \tilde{O}^{5 a\beta}|P_1\rangle=\tilde{\bf\varepsilon}^{5 a}_{\mu\nu},
\ea
and introduce the variables
\ba
&&Q_1^{\alpha}(z)=q^{\alpha}+\frac{1}{2}P^{\alpha}_z,\nonumber\\
&&Q_2^{\alpha}(z)=q^{\alpha}-\frac{1}{2}P^{\alpha}_z,
\ea
where $(z)$ is now meant to denote both variables $(z_+,z_-)$. The presence of a new variable $Q_2$, compared 
to \cite{blum&rob}, is related to the fact that we are parameterizing each single bilinear 
covariant rather then linear combinations of them, as in the electromagnetic case.

After some re-arrangements we get
\ba
\label{equa1}
&&\tilde{S}^{a}_{\mu\nu}=g_{u}\int Dz\frac{F^{a(\nu)}(z)}{(Q_1^2+i\eps)}\left\{\left[-g_{\mu\nu}\ds{q}+q_{\nu}\gamma_{\mu}+q_{\mu}\gamma_{\nu}\right]+\left[{P_z}_{\mu}\gamma_{\nu}+{P_z}_{\nu}\gamma_{\mu}\right]\right.\nonumber\\
&&\;\;\;\;\;\;\;\;\;\;\;\;\;\;\;\;\;\;\;\;\;\;\;\;\;\;\;\;\;\;\;\;\;\;\;\left.-\frac{\ds{q}}{(Q_1^2+i\eps)}\left[{P_z}_{\mu}{P_z}_{\nu}+{P_z}_{\mu}q_{\nu}+{P_z}_{\nu}q_{\mu}- g_{\mu\nu}(P_z\cdot q)\right]\right\}+\nonumber\\
&&\;\;\;\;\;\;\;\;g_{d}\int Dz\frac{F^{a(\nu)}(z)}{(Q_2^2+i\eps)}\left\{\left[-g_{\mu\nu}\ds{q}+q_{\nu}\gamma_{\mu}+q_{\mu}\gamma_{\nu}\right]-\left[{P_z}_{\mu}\gamma_{\nu}+{P_z}_{\nu}\gamma_{\mu}\right]\right.\nonumber\\
&&\;\;\;\;\;\;\;\;\;\;\;\;\;\;\;\;\;\;\;\;\;\;\;\;\;\;\;\;\;\;\;\;\;\;\;\left.+\frac{\ds{q}}{(Q_2^2+i\eps)}\left[-{P_z}_{\mu}{P_z}_{\nu}+{P_z}_{\mu}q_{\nu}+{P_z}_{\nu}q_{\mu}- g_{\mu\nu}(P_z\cdot q)\right]\right\}+\nonumber\\
&&\;\;\;\;\;g_{u}\int Dz\frac{G^{a(\nu)}(z)}{(Q_1^2+i\eps)}\left\{\left[-g_{\mu\nu}\frac{i\sigma^{\alpha\beta}q_{\alpha}\Delta_{\beta}}{M}+q_{\nu}\frac{i\sigma^{\mu\beta}\Delta_{\beta}}{M}+q_{\mu}\frac{i\sigma^{\nu\beta}\Delta_{\beta}}{M}\right]+\left[{P_z}_{\mu}\frac{i\sigma^{\nu\beta}\Delta_{\beta}}{M}+{P_z}_{\nu}\frac{i\sigma^{\mu\beta}\Delta_{\beta}}{M}\right]\right.\nonumber\\
&&\;\;\;\;\;\;\;\;\;\;\;\;\;\;\;\;\;\;\;\;\;\;\;\;\;\;\;\;\;\;\;\;\;\;\;\left.-\frac{i\sigma^{\alpha\beta}q_{\alpha}\Delta_{\beta}}{M(Q_1^2+i\eps)}\left[{P_z}_{\mu}{P_z}_{\nu}+{P_z}_{\mu}q_{\nu}+{P_z}_{\nu}q_{\mu}- g_{\mu\nu}(P_z\cdot q)\right]\right\}+\nonumber\\
&&\;\;\;\;\;g_{d}\int Dz\frac{G^{a(\nu)}(z)}{(Q_2^2+i\eps)}\left\{\left[-g_{\mu\nu}\frac{i\sigma^{\alpha\beta}q_{\alpha}\Delta_{\beta}}{M}+q_{\nu}\frac{i\sigma^{\mu\beta}\Delta_{\beta}}{M}+q_{\mu}\frac{i\sigma^{\nu\beta}\Delta_{\beta}}{M}\right]-\left[{P_z}_{\mu}\frac{i\sigma^{\nu\beta}\Delta_{\beta}}{M}+{P_z}_{\nu}\frac{i\sigma^{\mu\beta}\Delta_{\beta}}{M}\right]\right.\nonumber\\
&&\;\;\;\;\;\;\;\;\;\;\;\;\;\;\;\;\;\;\;\;\;\;\;\;\;\;\;\;\;\;\;\;\;\;\;\left.+\frac{i\sigma^{\alpha\beta}q_{\alpha}\Delta_{\beta}}{M(Q_2^2+i\eps)}\left[-{P_z}_{\mu}{P_z}_{\nu}+{P_z}_{\mu}q_{\nu}+{P_z}_{\nu}q_{\mu}- g_{\mu\nu}(P_z\cdot q)\right]\right\},\,\nonumber\\
\ea
with an analogous expressions for $S^{5 a}_{\mu\nu}$, that we omit,
since it can be recovered by performing the substitutions 

\ba
\label{repla}
\gamma_{\mu}\rightarrow \gamma^{5}\gamma_{\mu} \hspace{.4cm}\sigma^{\mu\nu}\rightarrow \gamma^5 \sigma^{\mu\nu}, \nonumber \\
F^{a(\nu)}, G^{a(\nu)}\rightarrow F^{5 a(\nu) }, G^{5 a(\nu)}
\ea
in (\ref{equa1}).

Similarly, for ${\bf \varepsilon}^{a}_{\mu\nu}$ we get
\ba
\label{equa2}
&&{\bf \varepsilon}^{a}_{\mu\nu}=g_{u}\int Dz F^{a(\nu)}(z)\left\{ \frac{1}{(Q_1^2+i\eps)}\eps_{\mu\alpha\nu\beta}\left[q^{\alpha}\gamma^{\beta} -\frac{P_z^{\beta}q^{\alpha}\ds{q}}{(Q_1^2+i\eps)}\right]\right\}-\nonumber\\
&&\;\;\;\;\;\;\;\;\;\;g_{d}\int Dz F^{a(\nu)}(z)\left\{ \frac{1}{(Q_2^2+i\eps)}\eps_{\mu\alpha\nu\beta}\left[q^{\alpha}\gamma^{\beta} +\frac{P_z^{\beta}q^{\alpha}\ds{q}}{(Q_2^2+i\eps)}\right]\right\}+\nonumber\\
&&\;\;\;\;\;\;\;\;\;\;g_{u}\int Dz G^{a(\nu)}(z)\left\{ \frac{1}{(Q_1^2+i\eps)}\eps_{\mu\alpha\nu\beta}\left[q^{\alpha}\frac{i\sigma^{\beta\delta}\Delta_{\delta}}{M} -\frac{P_z^{\beta}q^{\alpha}(i\sigma^{\lambda\delta}q_{\lambda}\Delta_{\delta})}{M\,(Q_1^2+i\eps)}\right]\right\}-\nonumber\\
&&\;\;\;\;\;\;\;\;\;\;g_{d}\int Dz G^{a(\nu)}(z)\left\{ \frac{1}{(Q_2^2+i\eps)}\eps_{\mu\alpha\nu\beta}\left[q^{\alpha}\frac{i\sigma^{\beta\delta}\Delta_{\delta}}{M} +\frac{P_z^{\beta}q^{\alpha}(i\sigma^{\lambda\delta}q_{\lambda}\Delta_{\delta})}{M\,(Q_2^2+i\eps)}\right]\right\}\,.
\ea 
The expression of $\tilde{\bf \varepsilon}^{5 a}_{\mu\nu}$ 
can be obtained in a similar way. 

To compute the $T^{Z_{0}}_{\mu\nu}$ tensor 
we need to include the following operators, which are related to the previous ones by $g_u\,, g_d\rightarrow 1$
\ba
&&\int dx^4 \frac{e^{iq x}x^{\alpha}}{2\pi^2\left(x^2-i\eps\right)^2}\langle P_2|S_{\mu\alpha\nu\beta}O^{a\beta}|P_1\rangle=S^{a}_{\mu\nu},\nonumber\\
&&\int dx^4 \frac{e^{iq x}x^{\alpha}}{2\pi^2\left(x^2-i\eps\right)^2}\langle P_2|S_{\mu\alpha\nu\beta} \tilde{O}^{5 a\beta}|P_1\rangle=\tilde{S}^{5 a}_{\mu\nu},\nonumber\\
&&\int dx^4 \frac{e^{iq x}x^{\alpha}}{2\pi^2\left(x^2-i\eps\right)^2}\langle P_2|\eps_{\mu\alpha\nu\beta} \tilde{O}^{a\beta}|P_1\rangle=\tilde{{\bf \varepsilon}}^{a}_{\mu\nu},\nonumber\\
&&\int dx^4 \frac{e^{iq x}x^{\alpha}}{2\pi^2\left(x^2-i\eps\right)^2}\langle P_2|\eps_{\mu\alpha\nu\beta} O^{5 a\beta}|P_1\rangle={\bf\varepsilon}^{5 a}_{\mu\nu}.
\ea
In this case a simple manipulation of (\ref{equa1}) gives
\ba
\label{equaZ0}
&&S^{a}_{\mu\nu}=\int Dz\frac{F^{a(\nu)}(z)}{(Q_1^2+i\eps)}\left\{\left[-g_{\mu\nu}\ds{q}+q_{\nu}\gamma_{\mu}+q_{\mu}\gamma_{\nu}\right]+\left[{P_z}_{\mu}\gamma_{\nu}+{P_z}_{\nu}\gamma_{\mu}\right]\right.\nonumber\\
&&\;\;\;\;\;\;\;\;\;\;\;\;\;\;\;\;\;\;\;\;\;\;\;\;\;\;\;\;\;\;\;\;\;\;\;\left.-\frac{\ds{q}}{(Q_1^2+i\eps)}\left[{P_z}_{\mu}{P_z}_{\nu}+{P_z}_{\mu}q_{\nu}+{P_z}_{\nu}q_{\mu}- g_{\mu\nu}(P_z\cdot q)\right]\right\}-\nonumber\\
&&\;\;\;\;\;\;\;\;\int Dz\frac{F^{a(\nu)}(z)}{(Q_2^2+i\eps)}\left\{\left[-g_{\mu\nu}\ds{q}+q_{\nu}\gamma_{\mu}+q_{\mu}\gamma_{\nu}\right]-\left[{P_z}_{\mu}\gamma_{\nu}+{P_z}_{\nu}\gamma_{\mu}\right]\right.\nonumber\\
&&\;\;\;\;\;\;\;\;\;\;\;\;\;\;\;\;\;\;\;\;\;\;\;\;\;\;\;\;\;\;\;\;\;\;\;\left.+\frac{\ds{q}}{(Q_2^2+i\eps)}\left[-{P_z}_{\mu}{P_z}_{\nu}+{P_z}_{\mu}q_{\nu}+{P_z}_{\nu}q_{\mu}- g_{\mu\nu}(P_z\cdot q)\right]\right\}+\nonumber\\
&&\;\;\;\;\;\int Dz\frac{G^{a(\nu)}(z)}{(Q_1^2+i\eps)}\left\{\left[-g_{\mu\nu}\frac{i\sigma^{\alpha\beta}q_{\alpha}\Delta_{\beta}}{M}+q_{\nu}\frac{i\sigma^{\mu\beta}\Delta_{\beta}}{M}+q_{\mu}\frac{i\sigma^{\nu\beta}\Delta_{\beta}}{M}\right]+\left[{P_z}_{\mu}\frac{i\sigma^{\nu\beta}\Delta_{\beta}}{M}+{P_z}_{\nu}\frac{i\sigma^{\mu\beta}\Delta_{\beta}}{M}\right]\right.\nonumber\\
&&\;\;\;\;\;\;\;\;\;\;\;\;\;\;\;\;\;\;\;\;\;\;\;\;\;\;\;\;\;\;\;\;\;\;\;\left.-\frac{i\sigma^{\alpha\beta}q_{\alpha}\Delta_{\beta}}{M(Q_1^2+i\eps)}\left[{P_z}_{\mu}{P_z}_{\nu}+{P_z}_{\mu}q_{\nu}+{P_z}_{\nu}q_{\mu}- g_{\mu\nu}(P_z\cdot q)\right]\right\}-\nonumber\\
&&\;\;\;\;\;\int Dz\frac{G^{a(\nu)}(z)}{(Q_2^2+i\eps)}\left\{\left[-g_{\mu\nu}\frac{i\sigma^{\alpha\beta}q_{\alpha}\Delta_{\beta}}{M}+q_{\nu}\frac{i\sigma^{\mu\beta}\Delta_{\beta}}{M}+q_{\mu}\frac{i\sigma^{\nu\beta}\Delta_{\beta}}{M}\right]-\left[{P_z}_{\mu}\frac{i\sigma^{\nu\beta}\Delta_{\beta}}{M}+{P_z}_{\nu}\frac{i\sigma^{\mu\beta}\Delta_{\beta}}{M}\right]\right.\nonumber\\
&&\;\;\;\;\;\;\;\;\;\;\;\;\;\;\;\;\;\;\;\;\;\;\;\;\;\;\;\;\;\;\;\;\;\;\;\left.+\frac{i\sigma^{\alpha\beta}q_{\alpha}\Delta_{\beta}}{M(Q_2^2+i\eps)}\left[-{P_z}_{\mu}{P_z}_{\nu}+{P_z}_{\mu}q_{\nu}+{P_z}_{\nu}q_{\mu}- g_{\mu\nu}(P_z\cdot q)\right]\right\}\,.\nonumber\\
\ea
The expression of $\tilde{S}^{5 a}_{\mu\nu}$ is obtained from 
(\ref{equaZ0}) by the replacements (\ref{repla}).

For the $\tilde{\bf \varepsilon}^{a}_{\mu\nu}$ case, a re-arrangement of 
(\ref{equa2}) gives 
\ba
\label{equaZ02}
&&\tilde{\bf \varepsilon}^{a}_{\mu\nu}=\int Dz F^{a(\nu)}(z)\left\{ \frac{1}{(Q_1^2+i\eps)}\eps_{\mu\alpha\nu\beta}\left[q^{\alpha}\gamma^{\beta} -\frac{P_z^{\beta}q^{\alpha}\ds{q}}{(Q_1^2+i\eps)}\right]\right\}+\nonumber\\
&&\;\;\;\;\;\;\;\;\;\;\int Dz F^{a(\nu)}(z)\left\{ \frac{1}{(Q_2^2+i\eps)}\eps_{\mu\alpha\nu\beta}\left[q^{\alpha}\gamma^{\beta} +\frac{P_z^{\beta}q^{\alpha}\ds{q}}{(Q_2^2+i\eps)}\right]\right\}+\nonumber\\
&&\;\;\;\;\;\;\;\;\;\;\int Dz G^{a(\nu)}(z)\left\{ \frac{1}{(Q_1^2+i\eps)}\eps_{\mu\alpha\nu\beta}\left[q^{\alpha}\frac{i\sigma^{\beta\delta}\Delta_{\delta}}{M} -\frac{P_z^{\beta}q^{\alpha}(i\sigma^{\lambda\delta}q_{\lambda}\Delta_{\delta})}{M\,(Q_1^2+i\eps)}\right]\right\}+\nonumber\\
&&\;\;\;\;\;\;\;\;\;\;\int Dz G^{a(\nu)}(z)\left\{ \frac{1}{(Q_2^2+i\eps)}\eps_{\mu\alpha\nu\beta}\left[q^{\alpha}\frac{i\sigma^{\beta\delta}\Delta_{\delta}}{M} +\frac{P_z^{\beta}q^{\alpha}(i\sigma^{\lambda\delta}q_{\lambda}\Delta_{\delta})}{M\,(Q_2^2+i\eps)}\right]\right\}\,.
\ea      
Also in this case, the expression of the ${\bf \varepsilon}^{5 a}_{\mu\nu}$ tensor is obtained  by the replacements (\ref{repla}).

\section{The partonic interpretation}
At a first sight, 
the functions $F^{(\nu)}, G^{(\nu)}, F^{5(\nu)}, G^{5(\nu)}$ do not have a simple partonic interpretation. 
To progress in this direction it is useful to perform the expansions 
of the propagators  
\ba
&&\frac{1}{Q_1^2+i \eps}\approx \frac{1}{2(\bar{P}\cdot q)}\frac{1}{\left[z_+ -\xi +\eta z_- +i\eps\right]},\nonumber\\
&&\frac{1}{Q_2^2+i \eps}\approx -\frac{1}{2(\bar{P}\cdot q)}\frac{1}{\left[z_+ +\xi +\eta z_- -i\eps\right]}
\ea
which are valid only in the asymptotic limit. In this limit only 
the large kinematical invariants and their (finite) ratios are kept.
In this expansion the physical scaling variable $\xi$ appears quite naturally and one is led to 
introduce a new linear combination 
\ba
t=z_+ +\eta z_-\,,
\ea
to obtain  
\ba
&&\frac{1}{Q_1^2+i \eps}\approx \frac{1}{2(\bar{P}\cdot q)}\frac{1}{\left[t-\xi +i\eps\right]},\nonumber\\
&&\frac{1}{Q_2^2+i \eps}\approx -\frac{1}{2(\bar{P}\cdot q)}\frac{1}{\left[t+\xi -i\eps\right]}.
\ea
Analogously, we rewrite $P_z$ using the variables $\left\{t,z_-\right\}$
\ba
P_z=2\bar{P}t +\pi z_-\,, 
\ea 
in terms of a new 4-vector, denoted by $\pi$, 
which is a direct measure of the exchange of transverse momentum with respect 
to $\bar{P}$
\ba
\pi=\Delta+ 2 \xi \bar{P}\,.
\ea
This quantity is strictly related to $\Delta_\perp$, as given in (\ref{delta}). 
The dominant (large) components of the process 
are related to the collinear contributions, and in our calculation the contributions proportional to the vector $\pi$ will be neglected. This, of course, introduces 
a violation of the transversality of the process of $O(\Delta_\perp/2\bar{P}\cdot q)$. 

Adopting the new variables $\left\{t,z_-\right\}$ and the conjugate ones $\left\{\bar{P},\pi\right\}$, the relevant integrals that we need to ``reduce'' to a single (partonic) variable  
are contained in the expressions 
\ba
\label{integrals}
&&H_{Q_1}(\xi)=\int dz_+ dz_- \frac{H(z_+,z_-)}{(Q_1^2+i\eps)}=\frac{1}{2\bar{P}\cdot q}\int Dz \frac{H(t+\xi z_-,z_-)}{(t-\xi+i\eps)}\nonumber\\
&&H_{Q_1}^{\mu}(\xi)=\int dz_+ dz_- \frac{H(z_+,z_-)}{(Q_1^2+i\eps)}\left[2\bar{P}^{\mu}z_+ +\Delta^{\mu}z_-\right]=\frac{1}{2\bar{P}\cdot q}\int Dz \frac{H(t+\xi z_-,z_-)}{(t-\xi+i\eps)}\left[ 2\bar{P}^{\mu}t+\pi^{\mu}z_-\right]\nonumber\\
&&H_{Q_1}^{\mu\nu}(\xi)=\int dz_+ dz_-\frac{H(z_+,z_-)}{(Q_1^2 +i\eps)^2}\left[P_z^{\mu}P_z^{\nu}+q^{\mu}P_z^{\nu}+q^{\nu}P_z^{\mu}-g^{\mu\nu}q\cdot P_z\right]\nonumber\\
&&\hspace{1.3 cm}=\frac{1}{(2\bar{P}\cdot q)^2}\int Dz\frac{H(t+\xi z_-,z_-)}{(t-\xi+i\eps)^2}\left[4\bar{P}^{\mu}\bar{P}^{\nu}t^2 + \left(2q^{\mu}\bar{P}^{\nu}+2q^{\nu}\bar{P}^{\mu}\right)t-g^{\mu\nu}(q\cdot P_z)\right.\nonumber\\ 
&&\hspace{6 cm}\left.+\pi^{\mu}\pi^{\nu}z_-^{2}+\left(q^{\mu}\pi^{\nu}+q^{\nu}\pi^{\mu}\right)z_- + \left(2\bar{P}^{\mu}\pi^{\nu}+2\bar{P}^{\nu}\pi^{\mu}\right)t z_-\right]\,.\nonumber\\
\ea
Here $H(z_+,z_-)$ is a generic symbol for any of the functions. 
We have similar expressions for the integrals depending on the momenta $Q_2$. 
  
The integration over the $z_-$ variable in the integrals shown above 
is performed by introducing a suitable spectral representation of the function $H(t,+\xi z_-,z_-)$.
As shown in \cite{blum&rob}, 
we can classify these representations by the $n=0,1,...,$ powers of the variable $z_-$ , 
\ba
\hat{h}_{n}(t/\tau,\xi)=\int dz_- {z_-}^{n}\hat{h}(\frac{t}{\tau}+\xi z_-,z_-).\,
\ea  
With the help of this relation one obtains
\ba
\label{repres}
&&\hat{H}_{n}(t,\xi)=\frac{1}{t^n}\int dz_- {z_-}^n H(t+\xi z_-,z_-)=\frac{1}{t^n}\int^{1}_{0}\frac{d\tau}{\tau} \tau^n \,\hat{h}_{n}(t/\tau,\xi)\nonumber\\
&&\hspace{1.5 cm}=\int_{t}^{sign(t)}\frac{d\lambda}{\lambda}\lambda^{-n}\hat{h}_{n}(\lambda,\xi)\,.
\ea
The result of this manipulation is to generate single-valued distribution amplitudes from double-valued ones. 
In the single-valued distributions $\hat{h}_{n}(t,\xi)$ the new scaling variables $t$ and $\xi$
have a partonic interpretation. $\xi$ measures the asymmetry between the momenta of the initial and final states, while 
it can be checked that the support of the variable $t$ is the interval $[-1,1]$. The twist-2 part of the Compton amplitude is related only to the $n=0$ moment of $z_-$. 
Before performing the $z_-$ integration in each integral of Eq.~(\ref{integrals}) using Eq.~(\ref{repres})
 - a typical example is $H_{Q_1}^{\mu\nu}(\xi)$ -
we reduce such integrals to the sum of two terms using the identity
\ba
\label{purpose}
\int_{-1}^{1}dt\frac{t^m}{(t\pm\xi\mp i\eps)^2}\hat{H}_{n}(t,\xi)=\int_{-1}^{1}dt\frac{t^{m-1}}{(t\pm\xi\mp i\eps)}\left[\hat{H}_{n}(t,\xi)-\frac{1}{t^n}\hat{h}_{n}(t,\xi)\right]\,.
\ea

As shown in \cite{Geyer}, after the $z_-$ integration, the integrals in (\ref{integrals}) can be re-written in the form 
\ba
&&H_{Q_1}(\xi)=\frac{1}{2\bar{P}\cdot q}\int_{-1}^{1}dt\frac{\hat{H}_{0}(t,\xi)}{(t-\xi+i\eps)},\nonumber\\
&&H_{Q_1}^{\mu}(\xi)=\frac{2\bar{P}^{\mu}}{2\bar{P}\cdot q}\int_{-1}^{1}dt\frac{t\hat{H}_{0}(t,\xi)}{(t-\xi+i\eps)}+ O(\pi^{\mu}),\nonumber\\
&&H_{Q_1}^{\mu\nu}(\xi)=\frac{1}{(2\bar{P}\cdot q)^2}\int_{-1}^{1}dt\frac{\left[2\hat{H}_{0}(t,\xi)-\hat{h}_{0}(t,\xi)\right]}{(t-\xi+i\eps)}4\bar{P}^{\mu}\bar{P}^{\nu}t\nonumber\\
&&\hspace{1.5 cm}+\frac{1}{(2\bar{P}\cdot q)^2}\int_{-1}^{1}dt\frac{\left[\hat{H}_{0}(t,\xi)-\hat{h}_{0}(t,\xi)\right]}{(t-\xi+i\eps)}\left\{(2q^{\mu}\bar{P}^{\nu}+2q^{\nu}\bar{P}^{\mu}-g^{\mu\nu}2q\cdot \bar{P})\right\}+O(\pi^{\mu}\pi^{\nu}),\nonumber\\
\ea
where, again, we are neglecting contributions from the terms proportional 
to $\pi^\mu$, subleading in the deeply virtual limit. 
The quantities that actually have a strict partonic interpretation are the $\hat{h}^{a}_{0}(t,\xi)$ functions, as 
argued in ref.~\cite{Roba&Horej}.
The identification of the leading twist contributions is 
performed exactly as in \cite{blum&rob}. We use a suitable form of the 
polarization vectors (for the gauge bosons) 
to generate the helicity components of the amplitudes 
and perform the asymptotic (DVCS) limit in order to identify the 
leading terms. Terms of $O(1/\sqrt{2\bar{P}\cdot q})$ are suppressed 
and are not kept into account. Below we will show only 
the tensor structures which survive after this limit.

\section{Organizing the Compton amplitudes}

In order to give a more compact expression for the amplitudes of our processes we define
\ba
&&g^{T \mu\nu}=-g^{\mu\nu}+\frac{q^{\mu}\bar{P}^{\nu}}{(q\cdot \bar{P})}+\frac{q^{\nu}\bar{P}^{\mu}}{(q\cdot \bar{P})},\nonumber\\
&&\alpha(t)=\frac{g_{u}}{(t-\xi+i\eps)}-\frac{g_{d}}{(t+\xi-i\eps)},\nonumber\\
&&\beta(t)=\frac{g_{u}}{(t-\xi+i\eps)}+\frac{g_{d}}{(t+\xi-i\eps)}\,.\nonumber\\
\ea 
Calculating all the integrals in the Eqs. (\ref{equa1}), (\ref{equaZ0}) and 
(\ref{equa2}), (\ref{equaZ02}), we rewrite the expressions of the amplitudes as follows 
\ba
\label{T}
&&T^{W^{+}}_{\mu\nu}=i U_{u d}\overline{U}(P_2)\left[i\left(\tilde{S}^{u}_{\mu\nu}+ S^{5 u}_{\mu\nu}\right) + {\bf \varepsilon}^{u}_{\mu\nu}+\tilde{{\bf \varepsilon}}^{5 u}_{\mu\nu}-i\left(\tilde{S}^{d}_{\mu\nu}+ S^{5 d}_{\mu\nu}\right) -{\bf \varepsilon}^{d}_{\mu\nu}-\tilde{{\bf \varepsilon}}^{5 d}_{\mu\nu}\right]U(P_1)\,,\nonumber\\
&&T^{W^{-}}_{\mu\nu}=-i U_{d u}\overline{U}(P_2)\left[i\left(\tilde{S}^{u}_{\mu\nu}+ S^{5 u}_{\mu\nu}\right) + {\bf \varepsilon}^{u}_{\mu\nu}+\tilde{{\bf \varepsilon}}^{5 u}_{\mu\nu}-i\left(\tilde{S}^{d}_{\mu\nu}+ S^{5 d}_{\mu\nu}\right) -{\bf \varepsilon}^{d}_{\mu\nu}-\tilde{{\bf \varepsilon}}^{5 d}_{\mu\nu}\right]_{g_u \rightarrow g_d}U(P_1)\,,\nonumber\\
&&T^{Z_{0}}_{\mu\nu}= i\,\overline{U}(P_2)\left[g_u g_{u V}\left(S_{\mu\nu}^{u}-i{\bf \varepsilon}_{\mu\nu}^{5 u}\right) -g_u g_{u A}\left(\tilde{S}_{\mu\nu}^{5 u}-i\tilde{\bf \varepsilon}_{\mu\nu}^{u}\right)\right.\nonumber\\
&&\hspace{3.8 cm}\left.-g_d g_{d V}\left(S_{\mu\nu}^{d}-i{\bf \varepsilon}_{\mu\nu}^{5 d}\right)+g_d g_{d A}\left(\tilde{S}_{\mu\nu}^{5 d}-i\tilde{\bf \varepsilon}_{\mu\nu}^{d}\right)\right]U(P_1)\,,
\ea
where, suppressing all the subleading terms in the tensor structures, we get
\ba
\label{S}
&&\overline{U}(P_2)\tilde{S}^{a \mu\nu}U(P_1)=\int_{-1}^{1}dt\,\alpha(t)\frac{g^{T\mu\nu}}{2\bar{P}\cdot q}\left[\overline{U}(P_2)\ds{q}U(P_1) \hat{f}_{0}^{a}(t,\xi) +\overline{U}(P_2)(i\frac{\sigma^{\alpha\beta}q_{\alpha}\Delta_{\beta}}{M})U(P_1)\hat{g}_{0}^{a}(t,\xi)\right]+\nonumber\\
&&\hspace{3.4 cm}\int_{-1}^{1}dt\,\beta(t) \frac{\bar{P}^{\mu}\bar{P}^{\nu}}{(\bar{P}\cdot q)^2}\left[\overline{U}(P_2)\ds{q}U(P_1) t \hat{f}_{0}^{a}(t,\xi) +\overline{U}(P_2)(i\frac{\sigma^{\alpha\beta}q_{\alpha}\Delta_{\beta}}{M})U(P_1) t \hat{g}_{0}^{a}(t,\xi)\right]\,\nonumber\\
\ea

while for the ${\bf \varepsilon}^{a \mu\nu}$ expression we obtain
\ba
\label{E}
&&\overline{U}(P_2){\bf \varepsilon}^{a \mu\nu}U(P_1)=\eps^{\mu\alpha\nu\beta}\frac{2q_{\alpha}\bar{P}_{\beta}}{(2\bar{P}\cdot q)^2}\int^{1}_{-1}dt\,\beta(t)\left[\overline{U}(P_2)\ds{q}U(P_1)\hat{f}_{0}^{a}(t,\xi)\right.+\nonumber\\
&&\left.\hspace{8.1 cm}\overline{U}(P_2)(i\frac{\sigma^{\alpha\beta}q_{\alpha}\Delta_{\beta}}{M})U(P_1)\hat{g}_{0}^{a}(t,\xi)\right]\,.
\ea
Passing to the $S^{a \mu\nu}$ and $\tilde{\bf \varepsilon}^{a \mu\nu}$ tensors, which appear in the $Z_0$ neutral current exchange, we get the following formulas 
\ba
\label{SP}
&&\overline{U}(P_2)S^{a \mu\nu}U(P_1)=\int_{-1}^{1}dt\left(\frac{1}{t-\xi+i\eps}+\frac{1}{t+\xi-i\eps}\right)\times\nonumber\\
&&\left\{\frac{g^{T\mu\nu}}{2\bar{P}\cdot q}\left[\overline{U}(P_2)\ds{q}U(P_1) \hat{f}_{0}^{a}(t,\xi)+\overline{U}(P_2)(i\frac{\sigma^{\alpha\beta}q_{\alpha}\Delta_{\beta}}{M})U(P_1)\hat{g}_{0}^{a}(t,\xi)\right]\right\}+\nonumber\\
&&\int_{-1}^{1}dt\left(\frac{1}{t-\xi+i\eps}-\frac{1}{t+\xi-i\eps}\right)\cdot
\frac{\bar{P}^{\mu}\bar{P}^{\nu}}{(\bar{P}\cdot q)^2}\left[\overline{U}(P_2)\ds{q}U(P_1) t \hat{f}_{0}^{a}(t,\xi) +\overline{U}(P_2)(i\frac{\sigma^{\alpha\beta}q_{\alpha}\Delta_{\beta}}{M})U(P_1) t \hat{g}_{0}^{a}(t,\xi)\right]\nonumber\\
\ea
\ba
\label{EP}
&&\overline{U}(P_2)\tilde{\bf \varepsilon}^{a \mu\nu}U(P_1)=\int_{-1}^{1}dt\left(\frac{1}{t-\xi+i\eps}-\frac{1}{t+\xi-i\eps}\right)\eps^{\mu\alpha\nu\beta}\frac{2q_{\alpha}\bar{P}_{\beta}}{(2\bar{P}\cdot q)^2}\times\nonumber\\
&&\hspace{4 cm}\left\{\left[\overline{U}(P_2)\ds{q}U(P_1)\hat{f}_{0}^{a}(t,\xi)+\overline{U}(P_2)(i\frac{\sigma^{\alpha\beta}q_{\alpha}\Delta_{\beta}}{M})U(P_1)\hat{g}_{0}^{a}(t,\xi)\right]\right\}\,.\nonumber\\
\ea
Obviously the $\tilde{S}^{a\,5 \mu\nu}$, $S^{a\,5 \mu\nu}$, $\tilde{{\bf \varepsilon}}^{a\, 5\, \mu\nu}$ and ${\bf \varepsilon}^{a\, 5\, \mu\nu}$ expressions are obtained by the substitution (\ref{repla}).
 
At this stage, to square the amplitude, we need to calculate the following quantity, separately for the two charged processes 
\ba
T^2= |T_{DVNS}|^2 + T_{DVNS}T_{BH}^* +T_{BH}T_{DVNS}^{*}+|T_{BH}|^2\,, 
\ea
which simplifies in the neutral case, since it reduces 
$|T_{DVNS}|^2$ \cite{ACG}.
Eqs.~(\ref{T})-(\ref{EP}) and their axial counterparts 
are our final result and provide 
a description of the deeply virtual amplitude in the electroweak 
sector for charged and neutral currents. The result can be expressed 
in terms of a small set of parton distribution functions 
which can be easily related to generalized parton distributions, 
as in standard DVCS. 

\section{Conclusions}
We have presented an application/extension of a method, which has been formulated 
in the past for the identification of the leading twist contributions to the parton amplitude in the generalized Bjorken region, 
to the electroweak case. We have considered the special case of a deeply virtual kinematics.
We have focused our attention on processes initiated by neutrinos. 
From the theoretical and experimental viewpoints 
the study of these processes is of interest,
since very little is known of the neutrino interaction at intermediate energy 
in these more complex kinematical domains.

\section{Acknowledgments}
We thank the Theoretical Physics Group at the Univ. of Ioannina, Greece, and in 
particular G. Leontaris, K. Tamvakis, J. Vergados and L. Perivolaropoulos 
for the warm hospitality provided while completing this work and for related discussions. 
This work is partly supported by the Pythagoras-1 Program (EPEAEK) 
of the Greek Ministry of Education and by INFN-BA21.


\normalsize

\begin{thebibliography}{99}

\bibitem{Marciano1} M.V. Diwan et al., Phys. Rev. D {\bf 68} 012002,2003; W.T. Weng et al, J. Phys. G {\bf 29} 1735, (2003); 
W. Marciano, talk at the KITP Conference:{\em  Neutrinos: Data, Cosmos, and Planck Scale} (Mar 3-7, 2003)
\bibitem{Winter} K. Winter, {\em Neutrino Physics}, 2nd ed. Cambridge Univ. Press, Cambridge 2000. 
\bibitem{Mangano} M. Mangano et al., Report of the nuDIS Working Group for the ECFA-CERN Neutrino- Factory study,  hep-ph/0105155. 
\bibitem{Morfin} J.G. Morfin, M. Sakuda, Y, Suzuki, (ed.) Proc. of the 
1st Workshop on Neutrino-Nucleus interactions in the Few GeV Region 
(Nuint01), Nucl. Phys. B, Proc. Suppl. {\bf 112}, (2002).
\bibitem{nuint} F. Cavanna, C. Keppel, P. Lipari, M. Sakuda (ed.), Proceedings of the Intl. Workshop on Neutrino 
Nucleus interactions in the Few GeV Region (NuInt04), Nucl. Phys. B, Proc. Suppl. {\bf 139}, (2005). 

\bibitem{Quigg} R. Gandhi, C. Quigg, M.H. Reno ad I. Sarcevic, Phys. Rev. {\bf D58}:093009, (1998). 
\bibitem{CL} C. Corian\'o and H.N. Li, JHEP 9807:008, (1998); Nucl. Phys. {\bf B 434}, 535 (1995).
\bibitem{ACG} P. Amore, C. Corian\`{o} and M. Guzzi, hep-ph/0404121.
\bibitem{Pol_Weiss} M.V. Polyakov and C.Weiss Phys. Rev D {\bf 60} 114017 (1999)

\bibitem{blum&rob} J.Bl\"umlein and D.Robaschik Nuc. Phys B {\bf 581} 449 473 (2000)

\bibitem{Lazar} M. Lazar, hep-ph/0308049 Ph.D. Thesis.

\bibitem{lazar1} J.Bl\"umlein, B. Geyer, M. Lazar and D.Robaschik, Nucl. Phys. Proc.Suppl. {\bf 89} 155-161 (2000)

\bibitem{lazar2} J.Eilers, B.Geyer and M.Lazar, Phys.Rev.D {\bf 69} 034015 (2004) 

\bibitem{Ji1} X.~Ji, Phys. Rev. D {\bf 55} 7114 (1997)  

\bibitem{Rad3} A.~V.~Radyushkin, Phys.\ Rev.\ D {\bf 56}, 5524 (1997)

\bibitem{Geyer} J. Bl{\"u}mlein, B. Geyer, D. Robaschik, Phys. Rev. D {\bf 65} 054029 (2002)

\bibitem{Rad1} A.~V.~Radyushkin, Phys.\ Rev.\ D {\bf 59} 014030 (1999) 

\bibitem{Man1} L. Mankiewicz, G. Piller, A.V. Radyushkin, Eur. Phys. J. C {\bf 10} 307 (1999)   

\bibitem{Roba&Horej} D. Robaschik and J.Horejsi, Fortsch. Phys. {\bf 42}: 101 (1994) 


\end{thebibliography}
\end{document}